# How Do You Want that Insulator?

*Gregory A. Fiete*
*Department of Physics*
*University of Texas at Austin, TX USA*

A normal insulator is turned into an exotic topological insulator by tuning its elemental composition.

An expert chef can take ordinary foods and bring out extraordinary flavors and textures, usually through a combination of the right ingredients and exacting cooking techniques. If the electrical conductivity of a material can be thought of as the flavor of a dish, nature can serve up specialties, such as ceramic insulators that become superconductors when served cold and when the ingredients (the chemical components) are carefully tuned. Superconductivity is an example of an emergent quantum phenomenon—one that is created by the coordinated motion of many particles (*1*). On page 560 of this issue, Xu *et al*. (*2*) report how tuning the composition of normal insulators can turn them into topological insulators, which are another example of materials exhibiting emergent quantum phenomena (*3–5*). This result has important implications for the promise of topological insulators in lower-power devices that rely on electron spin rather than charge.

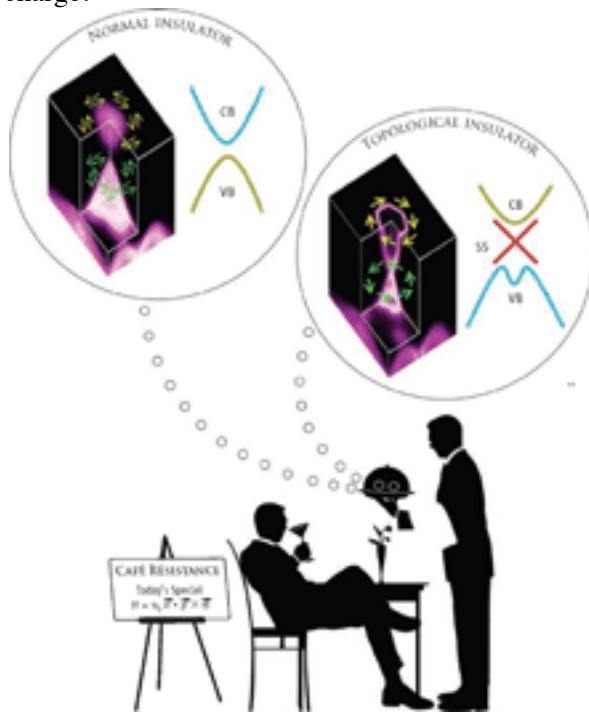

*Insulators made to order.*
*The choice at the Café Resistance is between an ordinary insulator on the left and the more exotic topological insulator on the right*. Xu et al. *show that both properties can be tuned into the material $BiTl(S_{1-\delta}Se_{\delta})_2$ by varying the fraction of sulfur and selenium. The normal insulator (sulfur-rich) has a band gap in the surface states, whereas the topological insulator (selenium-rich) has topologically protected metallic surface states running between the conduction and valence bands that cross at the Dirac point. The spin direction relative to the momentum changes sign for states above and below the Dirac point and is referred to as "texture inversion" by the authors*. CREDIT: DRAWINGS ADAPTED FROM ISTOCKPHOTO.COM

Most emergent quantum phenomena in condensed-matter physics require interactions among electrons, such as the pairing interaction between electrons in a superconductor. By contrast, the key ingredient for creating the topological insulator phase is spin-orbit coupling. Electrons have a quantum property known as spin that causes them to act in some respects like tiny bar magnets. Spin-orbit coupling forces a particular relation between the orientation of the spin and the orbital motion of the electron in space. In materials composed of light atoms, spin-orbit coupling effects are small and can be neglected, but when heavy atoms are present, their high nuclear charge can lead to appreciable magnetic fields in the frame of reference of the moving electron. This field couples to the spin of the electron and can drive a phase transition to the topological insulator state (*3*–*5*). Mathematically, the difference between a normal insulator and a topological one can be described with ideas borrowed from topology, hence the name topological insulator.

To appreciate the unusual nature of topological insulators, it is useful to see what causes a normal insulator (a ceramic) to be a poor conductor. The electron energy levels of a solid originate from those of the constituent atoms, and because the atoms are so numerous, they form dense bands of states: the lower-energy valence and the higher-energy conduction bands. In an insulator, the valence and conduction bands are separated by an energy gap, and the zero of the energy, known as the Fermi energy, lies in this gap where there are no conducting states (see the figure).
The unusual properties of topological insulators manifest at their surface, where additional states emerge that allow electrons to be mobile, while the bulk remains an insulator (*3*–*5*). The energy levels of these mobile surface electrons populate the band gap of the bulk material, and when plotted as a function of momentum, they form a constant-energy surface that is approximately circular (see the figure). The novel surface features derive from the composition of the bulk insulator and are not found in any other known system in nature.

Systematic and reliable control of the electrical properties of the surfaces of topological insulators, which is critical for applications, has presented a challenge (*6*). Xu *et al.* demonstrate a transition from a normal to a topological insulator by changing its bulk composition—replacing lighter sulfur (S) with heavier selenium (Se) to increase spin-orbit coupling. The material they studied, $BiTl(S_{1-\delta}Se_\delta)_2$ (where Bi is bismuth and Tl is thallium), starts to display a transition to topological insulator behavior at 40% Se, which fully develops by 60%.

To visualize this transition, the authors used spin- and angle-resolved photoemission spectroscopy to measure the spin and electron energy as a function of electron momentum. The replacement of S with Se creates metallic surface states with a linear energy-momentum relation. These surface extensions of the conduction and valence bands touch at a single node, called the Dirac point. The authors also observed subtle changes in the crystal lattice with x-ray scattering data. First-principles electronic structure calculations suggest that the transition depends on both the lattice changes and enhanced spin-orbit coupling.

Xu *et al.* take a further important step toward tuning the properties of the topological insulator by dosing a molecule, $NO_2$, on the surface that allows a tuning of the Fermi energy of the surface states. This step leads to a "texture inversion" of the spin-momentum relations when the Fermi energy passes through the Dirac point, as illustrated in the figure. Most of the novel electric and magnetic responses of topological insulators known to date rely on this tuning of the Fermi energy near the Dirac point (*7*, *8*). A theoretical proposal for a "topological excitonic condensate," an unusual symmetry-broken state with fractional charges (±e/2, where e is the charge of the electron) that could be formed by topological insulators, may also be a step closer with this new experimental technology (*9*).

Despite the experimental accomplishments reported by Xu *et al.*, there are many challenges that remain if topological insulators are to become functional components of electronic devices. Chief among them are the problems of "aging"—the material properties degrade on a time scale of hours to days. Also, the bulk conductivity is unacceptably high and greater than what theory has predicted. However, there are good reasons to hope for substantial improvements in sample quality. The material $Bi_2Te_2Se$ (where Tl is replaced by tellurium) was recently shown (*10*) to have a more insulating bulk, with up to 70% of the electrical conductance coming from the surface—more than two orders of magnitude better than most topological "insulators." Parallel methods of sample fabrication—chemical synthesis and molecular beam epitaxy— are expected to continue to lead to sample improvements in the near future.

Thus far, the known topological insulators are derived from materials with s- and p-type orbitals (which typically have weak electron correlations), and experiment has largely operated in the mode of confirming theoretical predictions. A new frontier with experimental surprises likely lies in the direction of more strongly correlated materials with d-and f-type electrons (*11*–*16*), which should expand the exciting choices already on the insulator menu.